\documentclass{aa}  

\usepackage{graphicx}
\usepackage{color}
\usepackage{subfigure}
\usepackage{ulem}
\usepackage{appendix}
\usepackage{inputenc}
\usepackage{txfonts}

\def\me{$\,{\rm M}_{\oplus}\,$}
\def\re{$\,{\rm R}_{\oplus}\,$}
\def\h2o{H$_2$O}
\def\sio2{SiO$_2$}
\def\gc3{g\,cm$^{-3}$}
\def\swr{Schwarzschild }
\def\ldx{Ledoux }
\definecolor{gry}{gray}{0.5}

\begin{document}

   \title{Explaining the low luminosity of Uranus:\\
  A self-consistent thermal and structural evolution}
   
   \author{Allona Vazan
          \inst{1,2}
          \and
          Ravit Helled
          \inst{1} }
          
   \institute{ Institute for Computational Science, Center for Theoretical Astrophysics \& Cosmology University of Zurich\\
    Winterthurerstr. 190, CH-8057 Zurich, Switzerland.  \\
    \and
   Racah Institute of Physics, The Hebrew University, Jerusalem 91904, Israel \\     \email{allona.vazan@mail.huji.ac.il} 
     }

  \abstract
{The low luminosity of Uranus is a long-standing challenge in planetary science. 
Simple adiabatic models are inconsistent with the measured luminosity, which indicates that Uranus is non-adiabatic because it has thermal boundary layers and/or conductive regions. 
A gradual composition distribution acts as a thermal boundary to suppress convection and slow down the internal cooling.
Here we investigate whether composition gradients in the deep interior of Uranus can explain its low luminosity, the required composition gradient, and whether it is stable for convective mixing on a timescale of some billion years. 
We varied the primordial composition distribution and the initial energy budget of the planet, and chose the models that fit the currently measured properties (radius, luminosity, and moment of inertia) of Uranus.
We present several alternative non-adiabatic internal structures that fit the Uranus measurements.
We found that convective mixing is limited to the interior of Uranus, and a composition gradient is stable and sufficient to explain its current luminosity.  
As a result, the interior of Uranus might still be very hot, in spite of its low luminosity.
The stable composition gradient also indicates that the current internal structure of Uranus is similar to its primordial structure.
{Moreover, we suggest that the initial energy content of Uranus cannot be greater than 20\% of its formation (accretion) energy}.
We also find that an interior with a mixture of ice and rock, rather than separated ice and rock shells, is consistent with measurements, suggesting that Uranus might not be "differentiated". 
Our models can explain the luminosity of Uranus, and they are also consistent with its metal-rich atmosphere and with the predictions for the location where its magnetic field is generated.}

\maketitle

\section{Introduction}

Uranus and Neptune, the ice giants in the Solar System, are often considered as twin planets, mainly because of the similar measurements of mass, radius, magnetic field, and atmospheric metallicity \citep{guillgaut14,helledguill18}.
One fundamental difference between the planets is their luminosity:  while the luminosity of Neptune seems to be consistent with an adiabatic structure, that of  Uranus is significantly lower, indicating very low to zero intrinsic flux \citep{hubbard95,fortney11a,nettel13}.

The planetary luminosity is an outcome of the planetary cooling history. The low luminosity of Uranus indicates that it has either lost all its energy or that the energy is still captured inside.
If the former is the case, the interior is cold and in thermal equilibrium with the solar radiation. Then, it is unclear how Uranus became so cold while the other planets are still cooling. 
{Although accelerated cooling by ice condensation can partially explain the low luminosity of Uranus, it cannot provide a full solution for the  luminosity problem \citep{kurosaki17}.}
The other possibility is that a gradual composition distribution affects the heat transport and slows the cooling down \citep{podolak91,marley95,podolak95}. 
A gradual distribution of the heavy elements is consistent with the interior constraints of Uranus \citep{podolak00,helled11UN}. An outward decrease in mean molecular weight can suppress convection and act as a thermal boundary \citep{ledoux47}. Such a thermal boundary insulates the inner heat from the outer envelope, and therefore the luminosity is low \citep{podolak91,nettel16}.

A composition gradient in the ice giant is a natural outcome of formation models. During planetary growth, accreted solids evaporate in the gaseous envelope mainly through friction and liberation of gravitational energy. 
Recent works that studied the composition distribution during planet formation \citep{helledsteven17, lozovs17,brouwers18,boden18,valleta18} showed that the resulting structure is probably gradual and not a distinct core-envelope structure. 
A gradual distribution is found for formation locations with relatively low solid-surface densities, as expected for the ice giants  \citep{helledsteven17}

The key question is then whether a gradual composition distribution can actually exist (rather than be assumed) in the interior of Uranus today. For this, the gradient would need to fulfill the following three criteria: (i) it would need to be stable against convection and mixing throughout the evolution, 
(ii) it would need to create a sufficient thermal boundary to slow the cooling and reproduce the measured luminosity of Uranus, and 
(iii) it would need to be consistent with the measurements of the current radius and gravitational field of Uranus.
To answer this question, a non-adiabatic model is needed, where the thermal and structure history of the planet are considered self-consistently to determine the consequent evolution or structure.

In previous studies we developed a detailed non-adiabatic structure evolution model for gas giants \citep{vazan15,vazan16}. This model also includes the change in the interior structure in time by convective mixing. 
The evolution of ice giants may be different from that of the gas giants.
The more metal-rich interior, lower mass, and high atmospheric metallicity affect the thermodynamic properties and the heat transport mechanism in the interior. We therefore expand our model by using the evolution features of metal-rich planets \citep{vazan18a,vazan18c}. 
We then apply our method to Uranus, and investigate whether a composition gradient is consistent with the low luminosity and the other measurements. Because we follow the entire evolution of the Uranus interior in detail, the initial properties can be derived by its current state.

This paper is structured as follows: in Section~2 we describe the initial composition and properties  of the model (Sec.~2.1, 2.2), the thermal and structural evolution methods (2.3), the nature of non-adiabatic structure evolution (2.4), and the parameters we fit (2.5). In Section~3 we present several examples of valid models of Uranus (3.1), emphasize the importance of the non-adiabatic evolution (3.2), and summarize the properties of all valid models in our study (3.3). We discuss aspects of the new models, and draw our conclusions in Section~4.

\section{Model}

In order to model the thermal and structural evolution of Uranus, we combined our thermal evolution calculation of a metal-rich planet \citep{vazan18c} with our planetary evolution code \citep{vazan15,vazan16}. 
The model allows for heat transport by radiation, convection, and/or conduction depending on the local conditions at each time step. The interior structure evolves by convective mixing in convective regions. The structure and evolution equations are solved simultaneously on an adaptive mass-time mesh \citep{vazan15}. 

\subsection{Parameter space of the interior structure}\label{sec:strct}

The initial planetary structure is characterized by the distribution of the heavy  elements (rock and/or ice) in radius. We varied both the heavy-element mass fraction  distribution $Z(r)$ and the composition of the assumed heavy elements. 
The parameter space of the $Z(r)$ distribution includes composition gradients of various slopes, between the extreme distinct core-envelope structure (two layers) and the shallowest gradient with an atmospheric enrichment of $Z<0.6$. We also considered  various slopes for the composition gradients on top of a distinct core.
The heavy-element distribution $Z(r)$ determines the initial metal enrichment in the outer gaseous atmosphere. 
Because hydrogen can be soluble in rock and ice at high pressures and temperatures  \citep{chattchen18}, we also considered cases with small fractions (up to 2\%) of hydrogen in the core. 

In most of our models the heavy elements are represented by a mixture of ice and rock.
The reason is that for the pressure conditions in the interior of Uranus we expect both ice and rock to be in ionic phases, and thus to favor a mixture \citep{hubbard95}. 
The ice-to-rock (i.e., water-to-rock) ratio in Uranus is unknown \citep{podolak12,helled11UN}.
We used a 2:1 ice-to-rock ratio, as is expected at the formation location of Uranus \citep[e.g.,][]{helledboden14}, as our standard case. However, we also  considered a Pluto-like ratio of more rock than ice (1:2), and cases with pure ice. 
For comparison, we also considered a three-layer model of a pure-ice shell on top of a rocky core.

We used the SCVH \citep{scvh} equation of state (EOS) for hydrogen and helium. In each mass layer the hydrogen and helium mass fractions are $X=0.72(1-Z)$ and $Y=0.28(1-Z),$ respectively, where $Z$ is the metal (rock+ice) fraction.
The EOS for ice (\h2o) and rock (\sio2) are improved versions of the calculation in \cite{vazan13}. 
The mixture of the heavy elements with hydrogen and helium was calculated using the additive volume law.
More details about the rock and ice EOS and their mixture can be found in Appendix~\ref{App:eos}.

\subsection{Interior energy content}\label{sec:enrg}

{The H-He mass in Uranus is much lower than the heavy-element mass \citep{helled11UN}, and therefore the heavy-element accretion energy dominates the overall primordial energy budget \citep{vazan18a}. 
The initial energy content of the envelope was not calculated explicitly because gas accretion processes (e.g., energy loss by accretion shock) are not significant in Uranus-mass planets \citep{marley07,mordasini17,Cumming2018}}.
The interior energy sources were modeled following \cite{vazan18c}, where the planetary energy was calculated accounting for the planetary formation (core accretion), iron differentiation, radioactive heating, solidification, and contraction. 

In this study we did not consider the energy associated with differentiation because the rock and the ice are assumed to remain mixed. 
The radioactive heating by the long-term radioactive elements (U, K, and Th) was taken as in \cite{nettel11} for the fraction of the rock in the ice+rock mixture. 
We also excluded latent heat of solidification because the solidification temperatures for a ice+rock mixture in high pressures is uncertain. Moreover, the temperatures in most of the ice+rock interior are usually above the critical point of both ice and rock (see Appendix~\ref{App:eos}).
The planetary contraction was automatically included through the hydrostatic structure and the mixture EOSs.

The initial energy content that is left in the metal-rich interior after its formation is a free parameter to fit the currently measured values of Uranus. 
{We estimated the formation energy using a simple description of the heavy-element accretion energy: the gravitational binding energy of the heavy-element-rich interior of mass $M$ and radius $R$ is $E_{binding}=3 G M^2/ 5 R$, where the maximum temperature $T_{max}$ corresponds to a case in which all the binding energy is converted into heat, that is, $E_{binding} = C_p M T_{max}$, where $C_p$ is the interior heat capacity.
We varied the initial energy budget by assuming different fractions of $E_{binding}$ as an initial condition for the long-term evolution \citep[see Section 2.1 in][for more details]{vazan18c}.}

\subsection{Evolution model - thermal and  structural}\label{sec:ev1}

The effect of the composition distribution on the heat transport was included in the thermal evolution, as in \cite{vazan15}. 
The heat transport was determined by the \ldx convection criterion \citep{ledoux47}, that is, convection takes place when and where $\nabla_R > \nabla_A + \nabla_X$. 
$\nabla_R$ and $\nabla_A$ are the radiative and adiabatic temperature gradients, respectively; and 
$\nabla_X=\sum_j\,[\partial \ln T(\rho,p,X)/{\partial X_j}]\,[{dX_j}/{d\ln p}]$
is the composition contribution to the temperature gradient, which depends on the mass fraction gradient (${dX_j}$) of each of the species ($j$). 
For a uniform composition $\nabla_X=0,$ and thus convection occurs when $\nabla_R>\nabla_A$, the simple convection criterion by \cite{schwarz06}.

Within a region that was found to be convective, we calculated the mixing of elements. The material transport by convective mixing was computed as a convective-diffusive process. The composition flux $F_j\propto D({\partial X_j}/{\partial m})$ was determined by the convective diffusion coefficient $D=a_0v_cl_c$. This coefficient depends on the convective velocity $v_c$ and the mixing length $l_c$, by a fraction between 0-1 ($a_0$). Here we took $a_0=0.1$.
The convective velocity was determined by the mixing parameter, $\alpha$, which is the ratio of the mixing length to the scale height, $\alpha=l_c/H_p$. Because the actual value is unknown, we considered $\alpha$ values between $5\times10^{-3}$ and 0.5 \citep[see the appendix in][for details]{vazan15}.
It should be noted that the scale height $H_p$ in the outer envelope of Uranus is on the order of the scale height of Jupiter. The higher mean molecular weight in the Uranus atmosphere combined with the lower gravity acceleration (Earth-like) causes the similarity of the mixing parameter of Uranus to that of Jupiter.

When the convection criterion is not fulfilled, the heat is transported by radiation (which dominates in low-density regions) or conduction (which dominates in high-density regions). 
Conduction and radiation were modeled as diffusive transports for a given opacity. 
The opacity was set by the harmonic mean of the conductive and radiative opacities.
The outermost envelope (optical depth lower than one) is the atmosphere. {The atmospheric} metallicity for the {radiaitive} opacity calculation was taken to be consistent with the initial envelope metallicity for each model. 
{We assumed a gray atmosphere (see Appendix A3 of \cite{vazan13} for details), and used} 
the analytical fit of \cite{valencia13} to the opacity tables of \cite{freedman08}. 
The atmospheric opacity, although uncertain, is crucial for determining the thermal evolution. Therefore we considered several other radiative opacity calculations, as described in Appendix~\ref{App:parameter}. 
{In addition to the uncertainty in the opacity calculation, the assumption of a gray atmosphere may also affect the planetary cooling history, and we hope to include a more realistic atmospheric models in future research.}
The conductive opacity was obtained as in \cite{vazan18c} to fit the conductivity of ice and rock in terrestrial conditions.
We used a constant albedo of A=0.3 and {an equilibrium temperature} of $T_{\rm eq}=59.1\pm1\,K$ \citep{guillgaut14}.

\subsection{Evolution of a non-adiabatic interior}\label{sec:ev2}

The non-adiabatic evolution allows for heat transport by convection, radiation, and/or conduction in each mass layer at each time step. 
{Thus, the interior is not necessarily adiabatic  during the evolution.}
As time progresses, the planet cools down from its surface through radiation. Below the radiative layer is a convective region whose thickness increases with time. The heavy-element distribution within the planet evolves as a result of convective mixing, when regions with composition gradients develop large-scale convection. 
The transition region between the outer convective envelope and the stable inner region with composition gradients is characterized by a discontinuity in composition and temperature.
When the (destabilizing) change in temperature at the transition dominates the (stabilizing) change in composition, the convective region progresses inward and the adiabatic region of the envelope expands. When the new composition discontinuity is sufficient to inhibit convection, the transition ceases to progress inward in mass.

In previous studies we showed that the interiors of Jupiter and Saturn can change significantly in time as a result of convective mixing when the composition gradient is shallow \citep{vazan16,vazan18b}.
A shallow composition gradient is not stable against convection, and convective mixing erases the gradient and leads to a homogeneous convective envelope. 
For Uranus, which is more metal rich and smaller, the composition gradient is limited to the a much smaller region than in the gas giants, and the composition gradient is therefore typically steeper. 

The heat in a stable (steep) composition gradient region is transported by conduction. The conductivity, although poorly constrained, is critical for properly simulating the planetary cooling. Moreover, the stable gradient, which is modeled as being conductive in our simulation, may develop layered convection, which has an intermediate heat-transport rate between large-scale convection and conduction (see Appendix~\ref{App:parameter} for further discussion).

However, if the gradient is too steep (stable), it may not be a sufficient thermal boundary to decelerate the cooling of Uranus and explain the luminosity that is measured. 
A rough estimate of the minimum thickness of the thermal boundary layer can be derived from the diffusion timescale $\tau_{\rm cond}=D^2\,\rho\,C_p/\kappa$. A layer of thickness $D$ is needed to slow the interior cooling for diffusive time of $\tau_{\rm cond}=5\times10^{9}$\,yr. 
For heat capacity $C_p\sim 1 kJ/kg/K$, density $\rho=1-5\times10^3 kg/m^3$, and thermal conductivity $\kappa=2-6 W/m/K$ \citep{stevenson83}, the thickness of the thermal boundary layer should be higher than several hundred kilometers in order to insulate the interior heat content. In the case of a more efficient heat transport, such as layered convection, the boundary length should be correspondingly higher.

\subsection{Fit to observations}\label{sec:fit}

The evolution models are constrained by measurements of the Uranus radius, effective temperature, and moment of inertia (MoI), where the mass, irradiation temperature, and albedo are input parameters. Ideally, the current-state structure models should be consistent with the measured $J_2$ and $J_4$, of Uranus, but the accuracy of the evolution model cannot be as high as the accuracies of the static interior structure models, and thus we fit MoI instead of the gravitational moments. 
We varied the $Z(r)$ distribution and composition, the initial (from formation) energy content, the radiative opacity, and the mixing length parameter, as described in Sec.~\ref{sec:strct}-\ref{sec:ev2}. 
Thermal evolution models are a step toward linking early (formation) stages and the current stage. Unlike static structure models, here the structure at a given time is the result of the evolution of the previous time step. The model parameters and their range (inputs and outputs) are summarized in Table~\ref{table1}.

\begin{table}[ht]
\centering
\begin{tabular}{ c c c c } \
\setlength{\tabcolsep}{0pt}
Parameter              & Value     &  Ref.   & Type\\
\hline
$M_p$  [\me]        &     14.539    &   (1)    & input\\ 
$T_{\rm irr}$  [K]   &     58.1     &   (1)    & input \\ 
$Albedo$          &     0.3         &   (1)    & input\\
$R_p$   [\re]      &   3.983-4.012   &   (1),(3)  & output \\
$L$ [erg/s]  &  0-7.2$\times10^{21}$ &  (1)  & output \\
$MoI$   [MR$^2$]     &   0.222-0.230    &  (2),(3)  & output\\
\hline
\end{tabular}
\caption{Model parameters: Parameters we used (inputs) and fit (outputs). References: (1) \cite{guillgaut14}, (2) \cite{nettel13}, (3) \cite{helled10}. }
\label{table1}
\end{table}

\section{Results}

\subsection{Uranus interior and evolution models}

We simulated the evolutionary tracks of hundreds of cases within the parameter space of our study (Sec.~\ref{sec:fit}). While most of the models failed to fit all Uranus measurements, several of them did fit, and they are presented in detail below. 

\begin{figure*}
\centerline{\includegraphics[width=20cm]{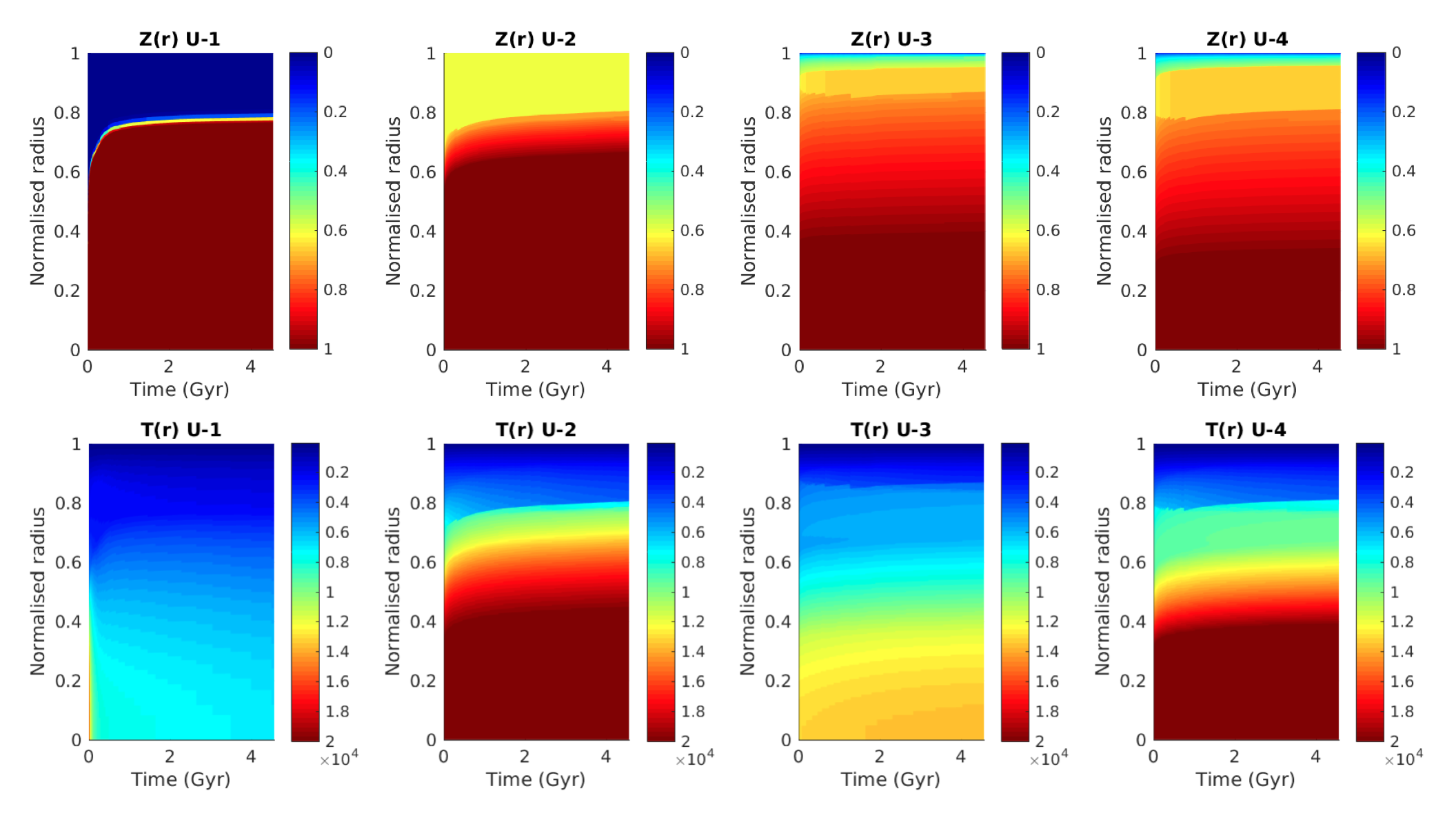}}
\caption{Thermal and structural evolution of Uranus (color) as a function of the radius layer (y-axis) and age (x-axis).
{\bf Upper panel:} Heavy-element mass fraction. {\bf Bottom panel:} Temperature profile.
The four cases are of valid Uranus models of different types: distinct layers (left), steep gradient (second), shallow gradient (third), and metal-rich shallow gradient (right).}\label{fig:ZT3d}
\end{figure*}

In Fig.~\ref{fig:ZT3d} we present the evolution of $Z(r)$ (up) and of the temperature profile (bottom) in Uranus for four representative models. All of these models are consistent with the observed parameters, but have different structures: a two-layer model, a steep gradient model, a shallow composition gradient model, and a steep rock-rich composition gradient. Within the parameter space we explored, more models fit the Uranus measurements, and the four models we show represent a given family of solutions. 
More details of the models appear in Table~\ref{table2}.

\begin{table*}
\centering
\begin{tabular}{l c c c c c c c} \
\setlength{\tabcolsep}{0pt}
Model number   & $R_p$ [\re] &  $L$ [erg/s]   & $MoI$ [MR$^2$]   & Z$_{env}$ & Z composition & Z$_{total}$ &  Initial energy (E$_{\rm acc}$)\\
\hline
U-1  & 3.994  & $6.24\cdot10^{21}$ & 0.222 & 0 & ice & 0.94 & 0.17 \\ 
U-2  & 3.989 & $5.46\cdot10^{21}$  & 0.229 & 0.60 & 2/3 ice, 1/3 rock & 0.93 & 0.15\\ 
U-3  & 4.012 & $1.99\cdot10^{21}$  & 0.230 & 0.11 & 2/3 ice, 1/3 rock & 0.88 & 0.09 \\ 
U-4  & 3.997 & $5.51\cdot10^{21}$ & 0.223 & 0.10 & 1/3 ice, 2/3 rock & 0.86 & 0.16 \\
\hline
\end{tabular}
\caption{Details of the Uranus models in Fig.~\ref{fig:Trho} at the current age: radius, luminosity, moment of inertia, outer envelope metallicity, metal composition, total mass of metals, and initial energy content (fraction of the total accretion energy: $E_{\rm acc}=3GM^2/5R$). The models are examples of different structure types for Uranus. }
\label{table2}
\end{table*}

The two-layer model (U-1), which has a distinct ice-envelope boundary, is the coldest. In this model each layer has a uniform composition, and the planet cools down through  large-scale convection (adiabatic cooling). This model knows no thermal boundaries, and the interior must therefore be cold in order to be consistent with the low luminosity of Uranus. We find that the current U-1 interior is cold enough to become (partially) conductive, in agreement with \cite{podolak19}. Hotter interiors of the two-layer model cannot fit the current luminosity of Uranus.
Model U-2 has a gradual distribution of an ice+rock mixture (2:1) in the interior, from $Z=1$ in the center to $Z=0.6$ in the outer envelope. 
The thermal boundary caused by the composition gradient keeps the interior hot while the outer envelope is insulated from the hot deep interior.

The composition gradient in model U-3 is wider than that of U-2, that is, the gradual region starts deeper in the interior and decreases all the way to the surface.
As a result, the mass fraction of hydrogen and helium in the interior is higher. 
Therefore U-3 must be colder in order to fit the current Uranus radius. Because the temperatures are lower in model U-3, the wider (shallower) composition gradient is found to be sufficient to prevent large-scale convection. The inner gradual region  then acts as a thermal boundary and slows the interior cooling down. 
U-3 like models with a hotter interior result in vigorous convective mixing. In these cases, a new adiabatic and metal-rich region is developed instead of the thermal boundary, and the models fail to fit Uranus MoI, luminosity, and/or radius at the current age. 
Model U-4 has a gradual distribution of a rock-rich mixture of ice+rock (1:2). The high mean molecular weight of this mixture requires a hotter interior and/or a lower total $Z$ to fit the measurements. 
Thus, although convective mixing is less efficient for higher mean molecular weight $Z$  \citep{vazan15}, the hotter interior results in a similar magnitude of convective mixing.

\begin{figure}
\centerline{\includegraphics[width=9.6cm]{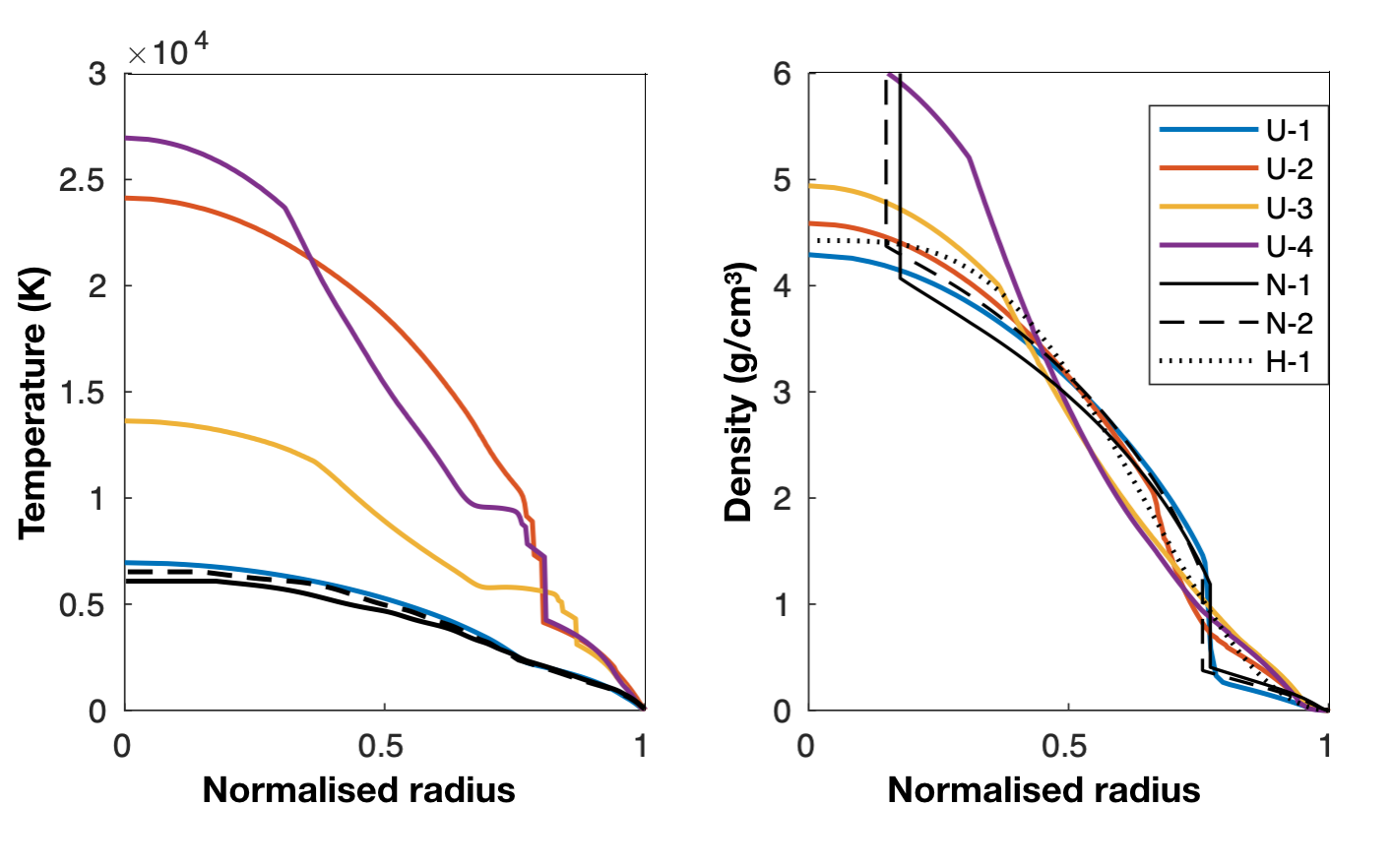}}  
\caption{Temperature (left) and density (right) profiles of the models of Fig.~\ref{fig:ZT3d} at the current Uranus age. For comparison, we show the Uranus model H-1 by \cite{helled11UN} and models N-1 and N-2 of \cite{nettel13}. {The kinks in the profiles} are caused by the non-adiabatic heat transport.}\label{fig:Trho}
\end{figure}

Next, we derived the interior properties of the present-day Uranus. In Fig.~\ref{fig:Trho} we show temperature (left) and density (right) profiles of Uranus for the four models presented in Fig.~\ref{fig:ZT3d}. We compare our results with the Uranus models of \cite{nettel13}, and Uranus polynomial density profile of \cite{helled11UN}. 
We find that our two-layer model is very similar to the models by \cite{nettel13}. The difference in density in the center is a result of our two-layer assumption (Sec.~\ref{sec:strct}) and has a small effect on the observed properties of Uranus. 
Our gradual models, on the other hand, are more consistent with the polynomial density profile of \cite{helled11UN}. 
It is clear from the temperature profiles in Fig.~\ref{fig:Trho} that the gradual composition region acts as a thermal boundary, and as a result, the inner region in these cases is much hotter than the outer envelope.
As expected, rock-rich interiors and colder interiors (of the same composition) are denser. 

When we further follow the structure evolution for longer than 4.55 Gyr, we find that some of the valid  models reach stable interior structures and some are still evolving.
The $Z(r)$ distribution of U-2 is already in a stable stage and will remain the same in the future evolution (for $t>4.55\,Gyr$).
The composition distribution in models U-3 and U-4 continues to change in the following gigayears of evolution, as the outer convective region progresses inward.
Model U-1 (core-envelope) is completely stable.  

\begin{figure}
\centerline{\includegraphics[width=9.6cm]{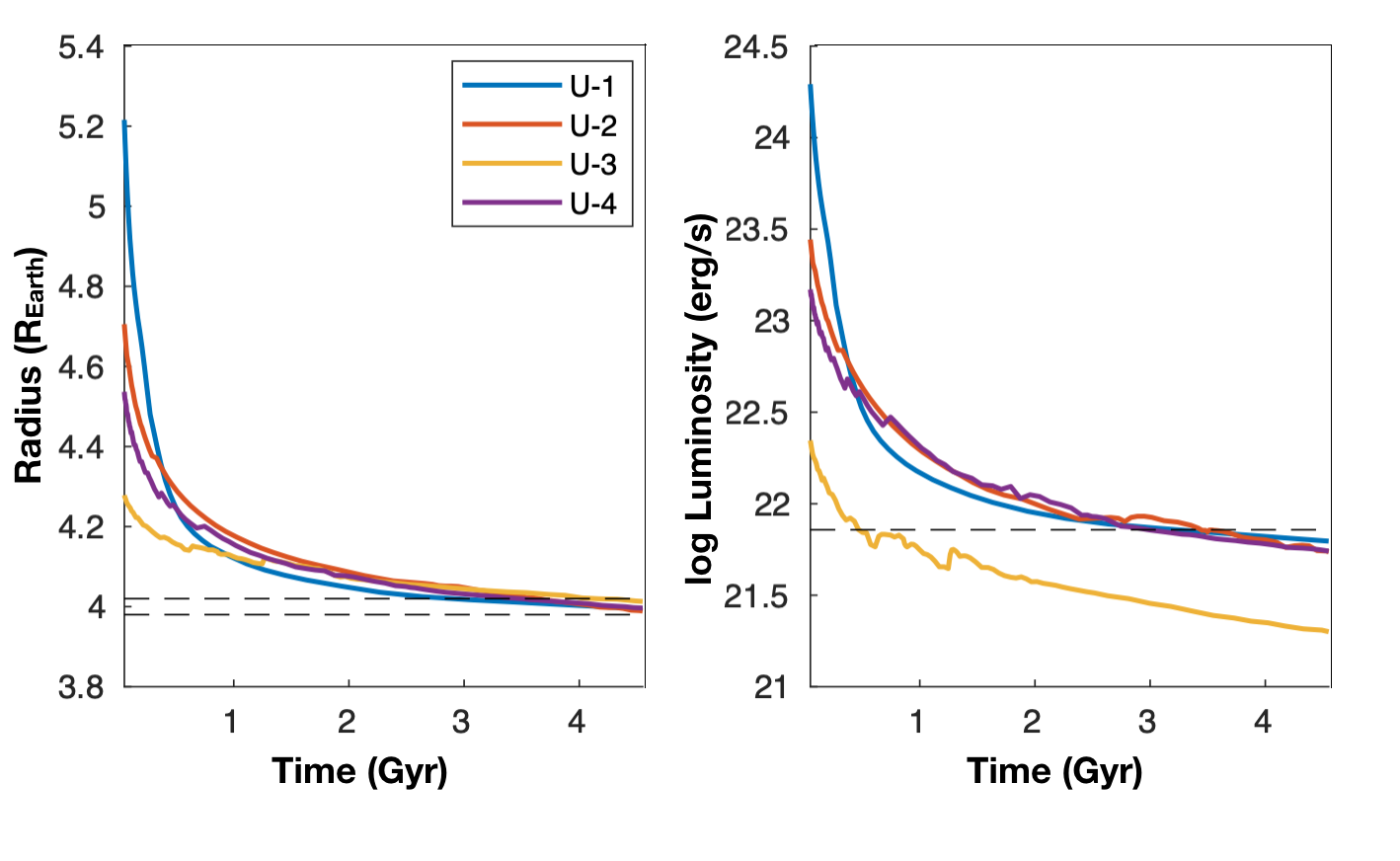}} 
\caption{ Radius (left) and luminosity (right) evolution for the four Uranus models  presented in Fig.~\ref{fig:ZT3d}. The horizontal dashed lines are for the measured radius (range) and luminosity (upper bound) of Uranus.
{Kinks in the profiles are due to} the non-uniform convective-mixing behavior.}\label{fig:RL}
\end{figure}

In Fig.~\ref{fig:RL} we show the radius and luminosity evolution  for the four models. Because the initial radius depends on the initial energy content, which is a free model parameter, the initial radii are not the same for all cases. 
The U-1 model, which has no thermal boundaries, cools down and contracts very fast to its current cold interior.
The other models are characterized by slower cooling, and their low luminosity is the result of the thermal boundary of the composition, which slows the heat transport from the inner interior to the surface. 
U-3 has an initially colder interior than U-2 and U-4, and therefore its luminosity is already lower in the early evolution stages. 
The profiles that are not smooth are the result of the non-uniform cooling in local regions in the interior, where convective mixing developed.

\subsection{Effects of non-adiabatic interior cooling}

\begin{figure}
\centerline{\includegraphics[width=9.4cm]{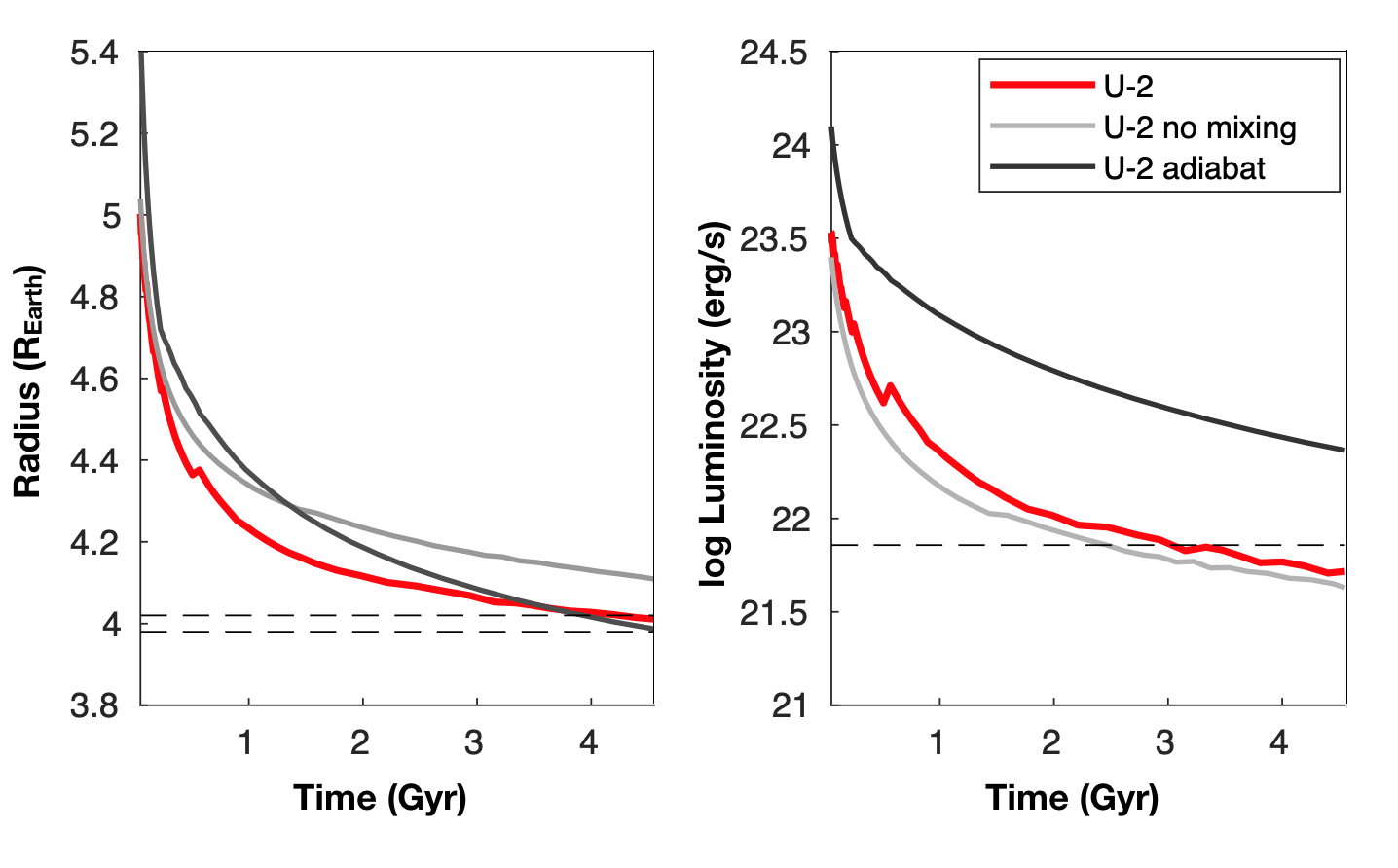}} 
\caption{Radius (left) and luminosity (right) of planets with identical structures. The standard model (red) is for the \ldx convection criterion and mixing in convective regions. Models without mixing (light gray) and without the composition effect on heat transport (dark gray) are shown. The horizontal dashed lines are for the measured radius (range) and luminosity (upper bound) of Uranus.
}\label{fig:led}
\end{figure}

The consequences of composition gradients on the non-adiabatic evolution is illustrated in Fig.~\ref{fig:led}. In this figure we recalculate the U-2 Uranus model, but once without considering convective-mixing and once without considering any effect of the composition gradient on the heat transport, that is, interior heat transport is similar to that of a fully convective planet. 
The figure shows that neither  model can fit the measured luminosity or radius of Uranus. The gray models are unphysical; they are shown here to emphasize the importance of heat transport and convective mixing in the presence of a non-uniform composition distribution. 

When convective mixing is ignored, we essentially force the structure to maintain its composition gradient, and by that slow down the heat transport and keep the outer envelope metal poor. The planetary contraction is thus decelerated, and the corresponding  radius is too large at the current age of Uranus. 
When we allow the entire planet to cool by large-scale convection, that is, when we ignore the thermal effect of the composition gradient, the planet cools down and contracts much faster, and we can reproduce the measured radius of Uranus. However, the luminosity of this adiabatic model is much too high. 
Overall, we find a 5-10\% change in radius by the non-adiabatic cooling of Uranus as a result of a composition gradient.
Thus, if the interior of Uranus consists of composition gradients, its  thermal evolution cannot be modeled by a simple large-scale convection model.

\subsection{Common properties of the valid Uranus models}

We find that the models that successfully explain the Uranus luminosity have some common properties: in these models the outer 20\% of the planetary radius develop a large-scale convection on top of a stratified inner region. This convective layer is metal rich for all models except for the two-layer model (U-1). Interestingly, this outer metal-rich convective region has {a heavy-element enrichment of 0.6-0.7 (mass fraction), and the water is in ionic phase for the current pressure-temperature regime (see water phases in Fig.~\ref{fig:red} below). This finding is} consistent with the prediction for the location where the magnetic field of Uranus is generated \citep{stanley04,stanley06}.

Considering the high metallicity of Uranus and the outer convective envelope, the composition gradients is rather steep. Because steep composition gradients are more stable against convection, most of the interior of Uranus remains stable. This result is consistent with the recent study of \citet{podolak19}. 
We also find that convective mixing, when it occurs, is limited to the outer part of the planet. Thus, unlike in the case of giant planets, the current structure of Uranus might not be very different from its primordial one. 

We also find that the temperatures in the deep interior are not well determined and can be significantly higher than the temperature expected from an adiabatic interior.
We find that the current central temperature of Uranus varies between 3000 and several tens of thousand (!) Kelvin.  
The deep interior might be insulated from the outer region, and therefore a very hot interior is possible. Such hot interiors are still consistent with the available measurements \citep{podolak19}. 
{The resulting hotter interiors may indicate different material properties and phases in the deep interior than found in standard Uranus structure models.  
An indication about the physical state of the water can be estimated from a comparison of the inferred thermodynamical conditions of Uranus with that of the water phase-diagram \citep[e.g.,][]{redmer11,wilson13}.

In Fig.~\ref{fig:red} we present the pressure-temperature history for models U-1 and U-2.
The profiles are overplotted on the water phase-diagram of \cite{redmer11}. As suggested by the figure, the pressure-temperature profiles of the non-adiabatic model (U-2) are clearly higher than in the adiabatic case (U-1).
The hotter interior is a result of less efficient heat transport from the deep interior, but can reproduce the observed current properties  of Uranus.
For the U-2 model, the water in the deep interior is expected to be in a plasma phase, while the water exterior to the steep composition gradient is in ionic phase. In the outer envelope (not shown here), the water is in a molecular phase. 
Because in most of our structure models the water and rock are mixed, the actual properties may differ from what appears in this diagram.}
Overall, the range of the pressure-temperature regime of Uranus in most of our models is much wider than in standard adiabatic models. 

\begin{figure}
\centerline{\includegraphics[width=9.4cm]{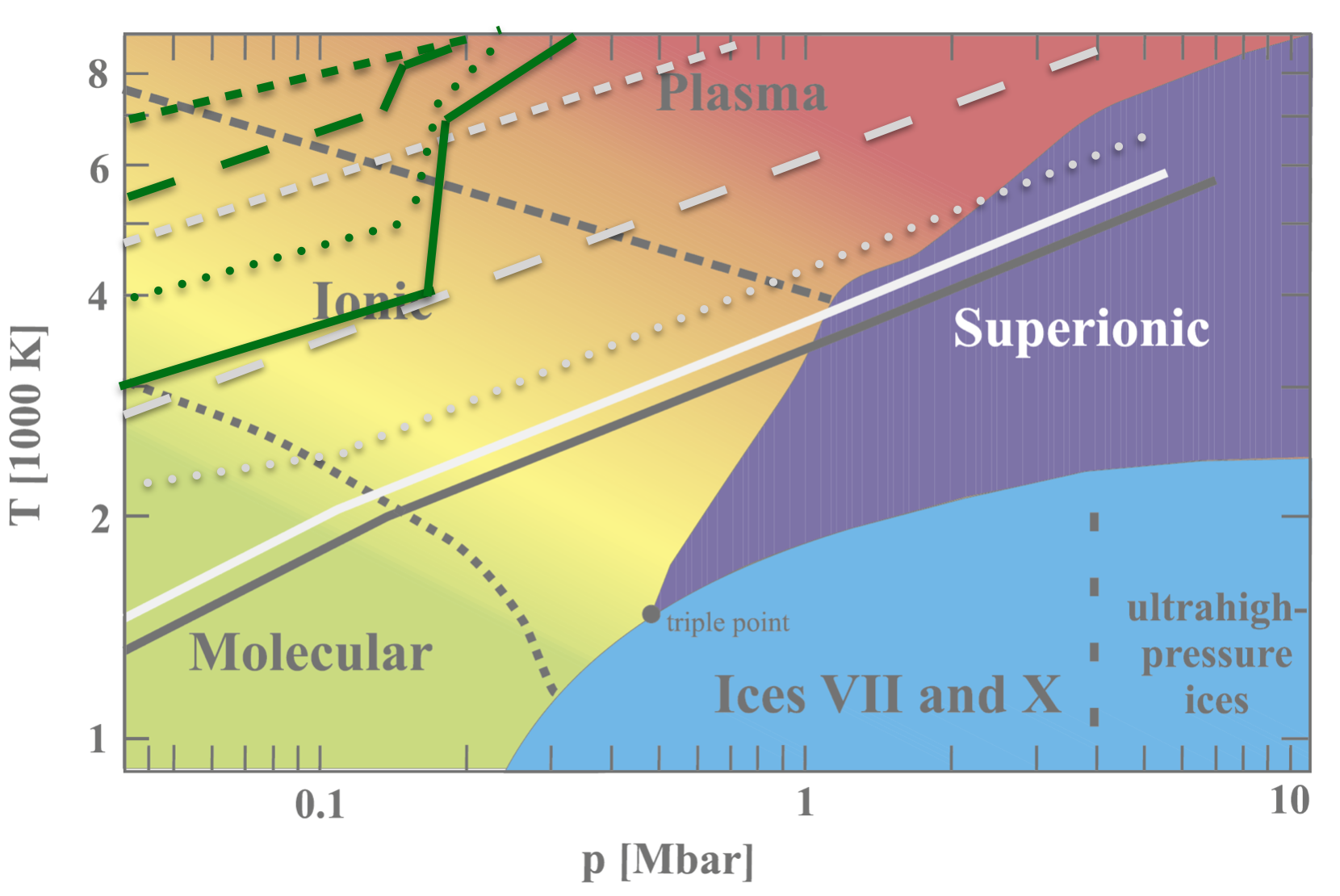}}
\caption{Pressure temperature profiles in time for models U-1 (gray) and U-2 (green) at ages of 10\,Myr (short dashed), 0.1\,Gyr (long dashed), 1\,Gyr (dotted), and 4.55\,Gyr (solid). 
The profiles are overplotted on the water phase-diagram of \cite{redmer11}. 
The adiabatic Uranus of \cite{redmer11} (solid gray) is similar to our U-1 model at present.
}\label{fig:red}
\end{figure}

The atmospheric opacity of all valid models does not include grains. 
High atmospheric opacity slows the planetary cooling down and usually results in a luminosity that is too high for the current state.
Models with grain opacity failed to fit the luminosity of Uranus, unless the initial interior is very cold. 
The total heavy-element mass in all our Uranus models varies within a range of 85\% to 95\% of the planet mass, in agreement with \cite{helled11UN} and \cite{podolak19}. The variation in heavy-element mass is by the heavy-element composition (rock-to-ice) and the interior temperature profile.
For example, for the same ice-to-rock ratio of 2:1, the hot gradual models are more metal rich (up to 95\%), while the cold gradual models contain only 85\%. As expected, the metallicity decreases when the rock enrichment is increased (1:2).

\section{Discussion}

\subsection{Required information for non-adiabatic models}\label{sec:req}

Our study demonstrates that the deep interior of Uranus could have a wide range of temperatures due to its insulation from the outer region. 
The temperatures in the planetary interior are of great importance for the thermodynamic state of the materials and their interactions \citep[e.g.,][]{keppler05,bali13,soubiran17}.
For example, the water-ice phases in Uranus are usually derived from adiabatic structure models \citep[e.g.,][]{redmer11,wilson13}, {while} the models we present here have much hotter interiors and the water is mostly in ionized and plasma phase {(see Fig.~\ref{fig:red})}. 
The different properties of water and rock in our model conditions (see Appendix~\ref{App:parameter} for discussion) have consequences on the thermodynamic processes that take place in the Uranus interior, and therefore on how the measurements are interpreted \citep[e.g.,][]{helled11UN, podolak12}.

The current available information for these properties is limited and therefore leads to a wide range of solutions. Future investigations of the material properties, both experimentally and theoretically, would allow us to narrow down the parameter space of possible solutions.

The pressure range in Uranus (and Neptune) varies between several bars to about 10\,Mbars. The temperature range inferred from our models is some hundred to several 10$^4$\,K. 
Within this pressure-temperature regime the properties of water, rock,  and their mixture is essential for calculating the evolution of the non-adiabatic interior. 
In particular, we need to improve our understanding of (1) the chemical interaction of rock and ice at high pressure, as well as the conductivity, phase transitions, and viscosity of the mixture. (2) We also still lack an understanding of the interaction of rock and ice with hydrogen at high pressures, (3) convection and mixing with the existence of composition gradients, and (4) the expected primordial composition of energy budget of young planets. 
In Appendix~\ref{App:parameter} we discuss some of the uncertainties in the thermal parameters (conductivity, radiative opacity, and viscosity) that can affect the heat transport, and in our model assumptions (layered convection and number of layers).

\subsection{Link to the initial energy budget}

The initial temperature profile used in evolution calculations is typically assumed to be  continuous across the core and the envelope. As a result, the core temperature is determined by the temperature at the bottom of the envelope, assuming an adiabatic structure. For standard initial envelope entropy of 7.5-9.5\,$K_b/baryon$ \citep[e.g.,][]{mordasini17}, the core of a Uranus-mass planet therefore contains 3\%-15\% of the solid accretion energy by the end of its formation. Moreover, \cite{vazan18c} showed that an initial energy content of more than 40\% of the accretion energy leads to envelope loss in metal-rich (90\% metals) core-envelope planets.

However, this assumption might not apply for non-adiabatic internal structures. Recent formation models indicate much higher initial interior temperatures by the gradual solid accretion \citep{lozovs17,brouwers18,boden18}. These models suggest that 10\%-40\% of the binding energy remains in the core after its formation.
Here we derive the initial energy budget directly from the solid accretion energy, as described in section~\ref{sec:enrg}, and not from a given formation model.
The initial energy content is therefore a free parameter that was varied to fit Uranus current measurements. 
This allowed us to limit the maximum energy content from the perspective of the current stage of Uranus, rather than from formation theory. In return, this constraint can then be used to guide formation models.

We found that if the primordial energy content is higher than 20\% of the gravitational binding energy, the measured properties of Uranus cannot be reproduced, {for both the adiabatic and non-adiabatic models. 
A core-envelope model cools down faster than the non-adiabatic models. Yet, its initial energy content must be limited in order to fit Uranus' observed properties. For gradual composition models, the primordial energy budget can be higher, but models with an initial energy content of more than 20\% still cannot reproduce the observed properties of Uranus.} 
In some of these models convection is too vigorous and changes the composition gradients, and in others the composition gradient is insufficient to delay the cooling.  
In all the cases where the initial energy budget is assumed to be greater than 20\% of the gravitational binding energy, the inferred radius and/or luminosity are higher than the measured radius and luminosity at the age of Uranus.  

Cold initial interiors with composition gradients can also fit observations, and therefore our limit on the initial energy content is an upper bound. A cold interior for Uranus could be an outcome of a giant impact followed by rapid cooling from the shock \citep{reinhardt19}.

Because the envelope of U-1 model is metal free and homogeneous, we compared its initial energy (entropy) with formation models \citep{mordasini17} and found a good agreement.
Although the envelopes of Uranus models with composition gradients are very hot, their   entropy is lower because of the strong effect of the heavy elements on the entropy, 
and therefore they cannot be directly compared with previous calculations.

It should be noted that the binding energy estimate we used here is highly simplified, and the primordial internal structure and thermal state are still poorly understood. 
In particular, layer convection (see Appendix B for details), which is not included in the our Uranus models, may change the requirement for the initial energy content.
Our simple estimate is a first step toward linking planetary origin with evolution and structure models. 
Clearly, the link to planet formation requires further investigation.  We suggest that this topic is addressed using a self-consistent calculation of the formation, evolution, and interior, and we hope to address this in future studies.

Moreover, we assumed that the composition gradients in Uranus are primordial and are determined by its formation process. The link to formation would no longer be valid if a giant impact occurred and significantly affected the energetics, as expected for Uranus. A giant impact is often required to explain the axis tilt and regular moons of Uranus \citep{stevenson86,podolak12},
and it might change the internal structure and energy profile  \citep{reinhardt19,liu19}. 
In a previous study we showed that gradients that form after the planet has cooled down significantly are more likely to survive because the internal temperature and thus the mixing efficiency is lower \citep{vazan16}. 
As a result, the primordial models assumed here could also represent Uranus shortly after a giant impact.

\section{Conclusions}

Our study shows that the low luminosity of Uranus can be explained by a structure (and evolution) with composition gradients. 
We calculated the thermal evolution of Uranus self-consistently with its structure evolution for a wide range of composition gradients. We find, rather than assumed, that a composition gradient between the metal-rich deep interior and the hydrogen-rich envelope is stable throughout several gigayears of evolution and is consistent with the observed properties of Uranus. 
Although a large portion of the Uranus interior can be convective, the intermediate gradual region isolates the inner hot region from the observed atmosphere.
The inefficient heat transport of a non-adiabatic interior suggests that several structure configurations are compatible with the measured properties of Uranus. Our main conclusions are summarized below.
\begin{enumerate}
    \item A composition gradient in the Uranus interior naturally  explains its low luminosity, without the need of artificial thermal boundaries. 
    Different types of composition gradients are stable during the evolution and are sufficient to slow down the cooling and fit the observed radius, moment of inertia, and luminosity.
    \item The initial energy content of Uranus cannot be greater than 20\% of its formation (solid accretion) energy. Primordial models with higher energy fail to fit the observations.
    \item A mixture of ice and rock in the deep interior of Uranus, rather than separate ice and rock shells, is consistent with the measured properties of Uranus, suggesting that Uranus might not be differentiated.
    \item Convective mixing is limited in the outermost region of Uranus. 
    This suggests that the current atmosphere of Uranus is similar to its primordial atmosphere, in contrast to the atmospheres of giant planets. 
    \item Two- and three-layer models of Uranus are able to fit Uranus properties only if the interior is very cold and (partially) conductive.
    \item Uranus is probably non-adiabatic, and therefore cannot be modeled by a simple large-scale convection model. The effect of the non-adiabatic cooling of Uranus (by composition gradient) on its current radius is 5-10\%.
    \item The total heavy-element mass fraction in Uranus is affected by the non-adiabatic evolution. The hot gradual models are more metal rich (up to 95\%) than the cold models ($\sim$85\%).
    \item The atmospheric opacity of Uranus cannot be very high during its evolution. Models with grain atmospheric opacity cannot explain the low  luminosity of Uranus unless the primordial interior is very cold. 

\end{enumerate}

While our work concentrated on Uranus, it can also be applied to Neptune.  
The fact that the luminosity of Neptune seems to be consistent with adiabatic cooling does not necessarily mean  that it is indeed adiabatic. 
Uranus and Neptune represent an important link in the chain between terrestrial planets and gas giants, and they are a key to understand planet formation. 
Nonetheless, the ice giants are the least-explored planets in our Solar System, and much is unknown about their interior properties.
It is therefore clear that efforts on both the modeling and observational fronts are needed. 
In addition, we suggest that in order to characterize the ice giants,  a better understanding of material properties and their interactions in high pressures and temperature conditions is needed.

Finally, a future space mission(s) to the ice giants might provide accurate measurements of their gravitational and magnetic fields and atmospheric compositions, which will then be used to further constrain their internal structure in its current state and will therefore improve our understanding of their origin and evolution.

\begin{acknowledgements}
  We like to thank the referee and the editor for constructive comments that improved the paper.
  R.H.~acknowledges support from the Swiss National Science Foundation (SNSF) Grant No. 200021\_169054. We acknowledge fruitful discussions within the ISSI "Formation of the Ice Giants" team meeting.
\end{acknowledgements}
 
\bibliographystyle{aa} 
\bibliography{allona}

\appendix
\section{Equation of state of ice and rock mixture}
\label{App:eos}

\begin{figure}[ht]
\centerline{\includegraphics[width=9.6cm]{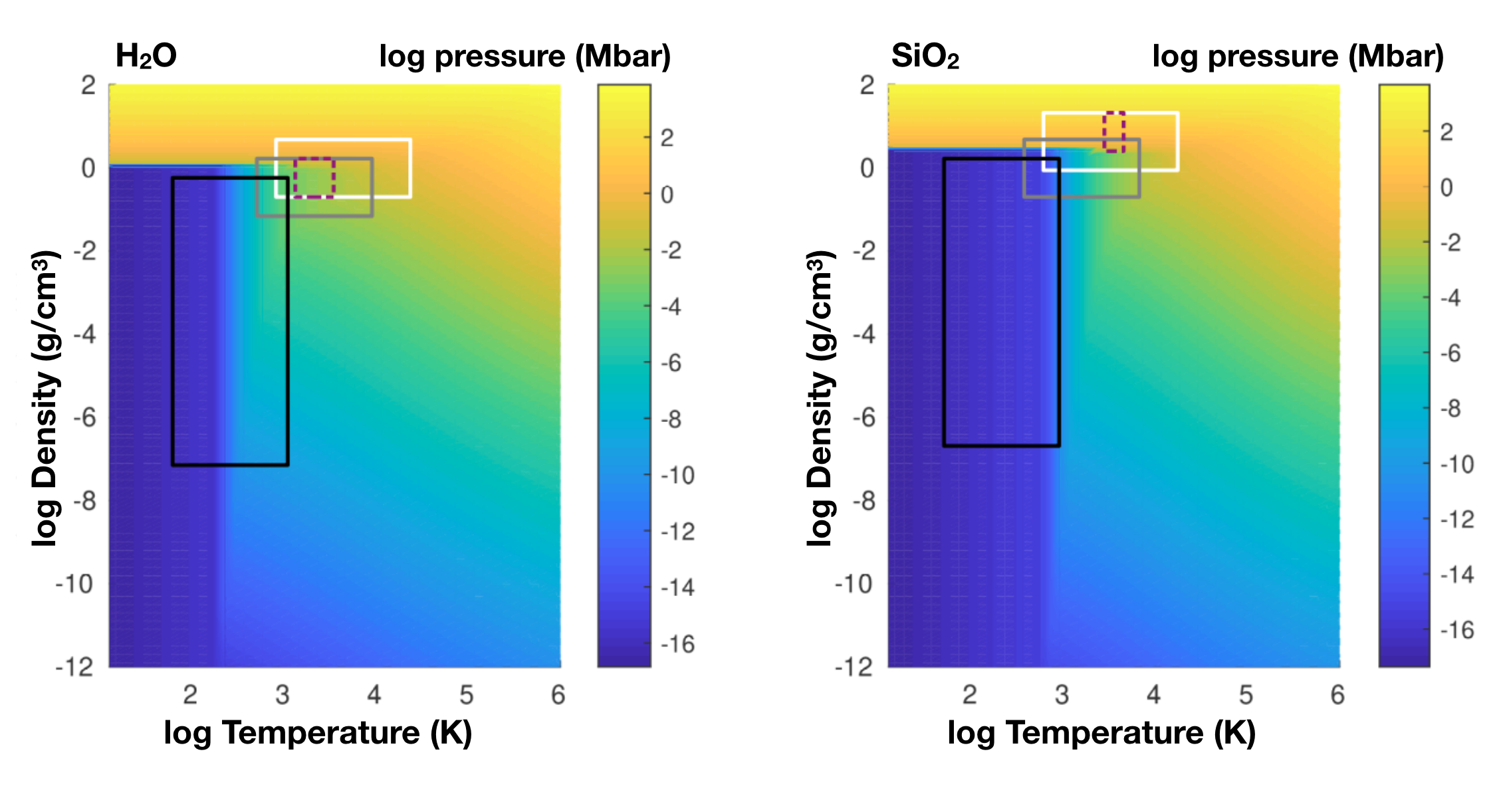}}
\caption{Our \h2o (left) and \sio2 (right) EOS: pressure (color) vs. temperature and density. Rectangles mark the relevant regimes of the Uranus inner Z region (white), gradient region (gray), and envelope (black), based on the models of this work. The regime for the adiabatic Uranus \citep{nettel13} is shown (dashed purple) for comparison. The EOS range for Uranus models is much larger than the range for the adiabatic structure. 
}\label{fig:eos}
\end{figure}

The equations of state (EOS) of ice (\h2o) and rock (\sio2) are improved versions of our calculation in \cite{vazan13}, based on the Quotidian equation of state (QEOS) method \citep{qeos88}.
The EOSs contain a solid or liquid phase and a gaseous phase, and cover a wide range of temperatures and densities. 
In Fig.~\ref{fig:eos} we show the pressure as a function of temperature and density of our ice and rock. The EOS regime that is relevant for the Uranus interior models that are presented in this work are marked with rectangles, as described in the figure caption.

In the new EOS version, the phase transition between the gaseous phase and the solid or liquid phase was calibrated by the room pressure-temperature point of evaporation.  
Therefore, the phase transition of the new version is more realistic.
For most of the temperature-density space, the new version is similar to the old version. 
The ice and rock EOSs are in good agreement with the EOS  of \cite{aneos72} and the EOS of \cite{sesame92}, as is shown in Fig.~3 in \cite{vazan13}.
The new water EOS by \cite{mazevet19} is denser than the others. 
For the modeling of the Uranus interior by an ice+rock mixture, the density effect is insignificant, however; it is equivalent to a slightly higher rock mass fraction in the ice+rock mixture. 

A mixture of ice and rock is expected at least in some of the interior because of fluid mineral interaction in high pressure \citep{keppler05}. 
The thermal properties for a mixture of ice and rock differ from the properties of the separate materials \citep[e.g.,][]{keppler05,kessel05,soubiran17}.
However, in our model we ignored chemical interactions between ice and rock for simplicity.
The ice and rock mixture in our model was calculated by the additive volume law: for a given pressure and temperature, the mixture density $\rho_Z(p,T)$ was calculated by $1/\rho_Z= Z_{H_2O}/\rho_{H_2O}+Z_{SiO_2}/\rho_{SiO_2}$.
Because the variables in the EOS tables are temperature and density, we iterated to solve for the mixture density. After the density of each species was found, the mixture energy and entropy were calculated accordingly \citep[see Appendix~A1 in][for details]{vazan13}.

\section{Uncertainties in heat transport}
\label{App:parameter}

When the thermal evolution is non-adiabatic, the local material properties significantly influence the thermal evolution. 
Because material properties in high pressure are rather uncertain, we discuss below the main uncertainties in the thermal parameters and assumptions of the model.

\underline{\textup{Conductivity}} 
In the absence of convection in the deep interior, heat is transported by conduction. The conductivity, although crucial for thermal evolution, 
is very uncertain in planetary conditions. For example, superionic water at high pressure has a different conductivity than low-pressure ice \citep[e.g.,][]{millot18}.
In addition, the conductivity in high temperatures, as in our models, behaves differently than in low temperatures \citep{vandenBerg10}. Above 5000 K, the thermal conductivity is strongly affected by the electronic contribution, which goes exponentially with temperature \citep{umemoto06,french19}.
Here we scaled the conductivity to fit the Earth values as described in \cite{vazan18c}.

\underline{Number of layers in the model}
The numerical nature of the model encapsulates a structure uncertainty when a gradual interior is modeled.
The numerical evolution of the structure is determined by the heat transport in each mass layer in the interior.
For each mass layer, the three temperature gradients (adiabatic, radiative, and composition) were calculated with respect to the neighbor layers, and the convection criterion was then tested.
The calculation requires constant properties (pressure, temperature, metallicity, etc.) within each mass layer.
Thus, if the composition distribution is gradual, the mass distribution per  shell affects the resulting temperature gradients and the thermal evolution.

In a previous work we tested this effect in the context of gas giants \citep[Sec~3.2.2 in][]{vazan15} and found it to have a slight effect on the result when the number of layers varied between 150 and 500.
The reason for the quite small effect is probably the flexible (adaptive) mass grid in our model.
The evolution model has an adaptive mass grid, that is, in a region with sharp changes in thermodynamic properties (pressure, temperature, or opacity), the resolution of the grid increases during the run. 
Here we used 500 mass grid points (layers) as our standard value. 
When we varied the number of layers between 100 and 500, we found a negligible difference in the current internal structure. We cannot exceed 500 layers due to a numerical difficulty.
The main difference we found is in the radius and luminosity fluctuations around the same general evolution slope. As expected, the fluctuations decrease with more layers.  
The final (current) radius and luminosity can vary by up to a few percent because of these fluctuations.

\underline{Radiative opacity} 
The radiative atmosphere is the outermost layer of the planet, which controls the planetary cooling and contraction rate. 
In non-adiabatic models, the radiative atmosphere strongly influences the long-term evolution, although it is not always the thermal bottleneck.
For interiors with composition gradients, the atmospheric opacity affects convective mixing, which depends on the cooling rate of the outer envelope.

\begin{figure}[ht]
\centerline{\includegraphics[width=9.6cm]{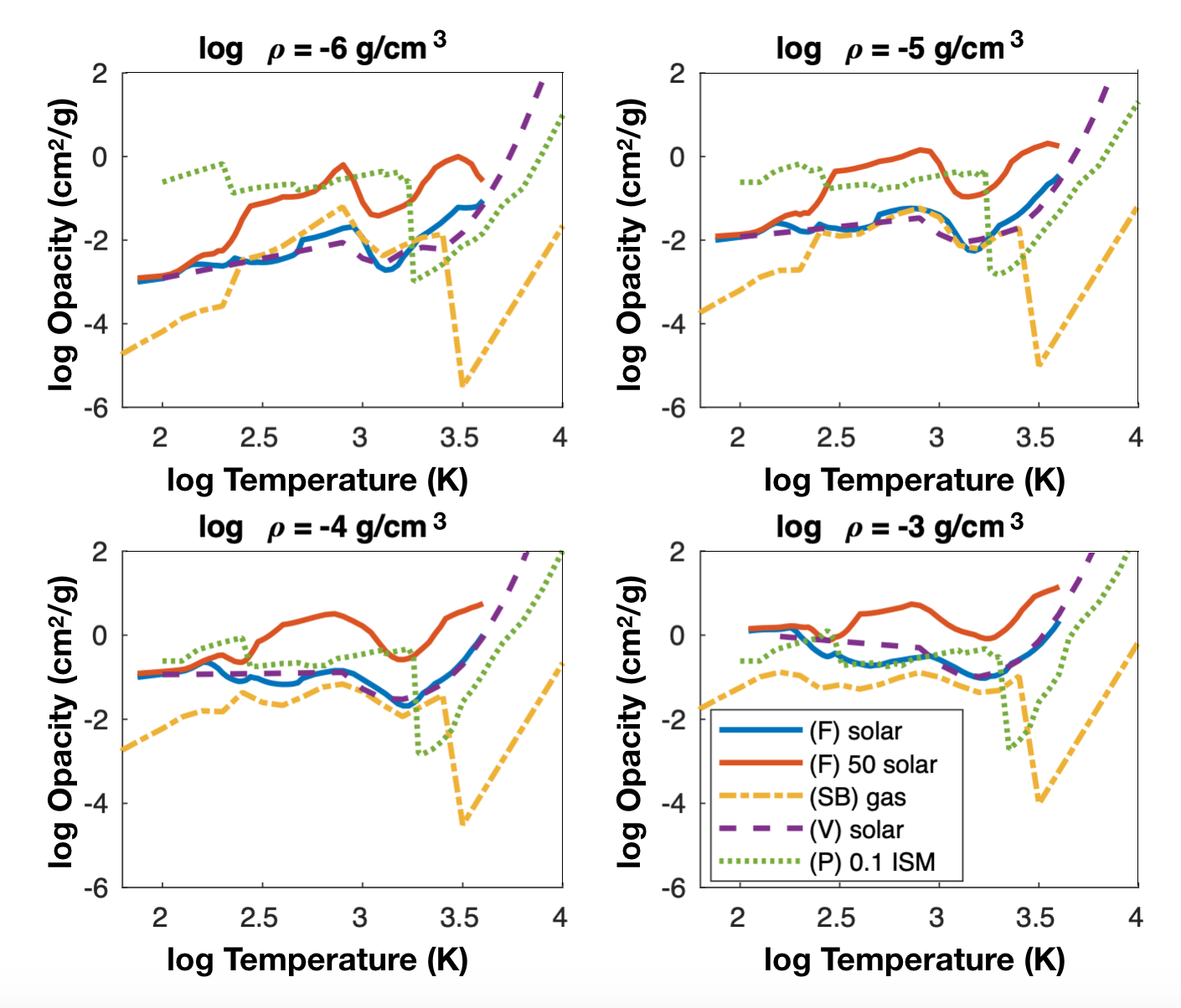}}
\caption{Atmospheric opacity calculated by various groups: (F) \cite{freedman08}, (SB) \cite{sharp07}, (V) \cite{valencia13}, and (P) \cite{pollack85}. The density range is for pressures between 0.1 millibar and 100 bar.
}\label{fig:opac}
\end{figure}

We here mainly used the opacity calculation of \cite{valencia13}, which allows for easy modification of the envelope metallicity. 
However, because radiative opacity is important for the thermal and structural evolution, we tested several other calculations. 
In Fig.~\ref{fig:opac} we show the radiative opacity as calculated by different groups for a given density. 
The density range fits pressures between 0.1 millibar to 100 bar. 
We present the \cite{freedman08} calculation for solar and for 50 times solar metallicity, the \cite{sharp07} calculation for gas, the \cite{valencia13} analytical fit for \cite{freedman08}, and one-tenth of the interstellar matter opacity of \cite{pollack85}. 
The figure shows that the differences in atmospheric opacities are rather large. 
This uncertainty leads to different measured properties for the same planet after several gigayears of evolution.

\underline{Viscosity} In a heavy-element-rich interior the viscosity determines the convective velocity \citep[e.g.,][]{stevenson83}.
The viscosity at high pressure depends on the material phase, which is difficult to determine at high pressure-temperature conditions. 
As is shown in Fig.~\ref{fig:eos}, the temperatures of the inner pure-Z region and most of the region with composition gradient are much higher than the critical ice+rock point \citep[e.g.,][]{kessel05}, and thy are even higher than the uncertain \sio2 dissociation conditions \citep{melosh07}.
Only the planetary envelope (black rectangle) is below the critical point of rock, and only part of it is below the critical point of ice.
Therefore we assume low (liquid) viscosity in the deep interior, which is also consistent with the viscosity calculation for ice \citep{french19}.

\underline{Layered convection} 
In planetary interiors with composition gradients, heat can be transported by layered or double-diffusive convection \citep{lecontechab12}. 
Layered convection occurs in regions that are found to be stable against convection according to the \ldx criterion, but unstable according to \swr criterion \citep{rosenblum11,wood13}, that is, in locations where the composition gradient suppresses large-scale convection. 
In our model we assumed that non-convective regions are conductive and/or radiative.
Layered convection, which is an intermediate heat transport mechanism, is not considered.
Therefore the heat transport rate in our model can be taken as a lower bound. 
As a result, our models provide an upper bound for the possible thermal boundary and the maximum effect of the composition gradient on the thermal evolution of Uranus.

\end{document}